\documentclass[a4paper,fleqn,usenatbib,twocolumn]{mnras}

\usepackage{newtxtext,newtxmath}

\usepackage[T1]{fontenc}
\usepackage{ae,aecompl}

\usepackage{graphicx}	
\usepackage{amsmath}	
\usepackage{amssymb}	
\usepackage[usenames,dvipsnames]{color} 

\definecolor{purple}{rgb}{0.5,0,0.5}

\title[Thermal Regulation and Star Formation]{Thermal Regulation and the Star-Forming Main Sequence}
\author[Charles L. Steinhardt et al.]{
Charles L. Steinhardt,$^{1,2}$\thanks{E-mail: steinhardt@nbi.ku.dk}
and Adam S. Jermyn,$^{3,4}$
Jackie Lodman, $^{5,1,2}$
\\
$^{1}$Cosmic Dawn Center (DAWN)\\
$^{2}$Niels Bohr Institute, University of Copenhagen, Lyngbyvej 2, DK-2100 Copenhagen \O \\
$^{3}$Center for Computational Astrophysics, Flatiron Institute,
New York, NY 10010, USA \\
$^{4}$Kavli Institute for Theoretical Physics, University of California at Santa Barbara, Santa Barbara, CA 93106, USA \\
$^{5}$California Institute of Technology, 1200 E California Blvd, Pasadena, CA 91125, USA }

\date{Accepted XXX. Received YYY; in original form ZZZ}

\pubyear{2019}

\begin{document}
\label{firstpage}
\pagerange{\pageref{firstpage}--\pageref{lastpage}}
\maketitle
\begin{abstract}
We argue that the interplay between cosmic rays, the initial mass function, and star formation plays a crucial role in regulating the star-forming "main sequence".  To explore these phenomena we develop a toy model for galaxy evolution in which star formation is regulated by a combination of a temperature-dependent initial mass function and heating due to starlight, cosmic rays and, at very high redshift, the cosmic microwave background. This produces an attractor, near-equilibrium solution which is consistent with observations of the star-forming main sequence over a broad redshift range. Additional solutions to the same equations may correspond to other observed phases of galaxy evolution including quiescent galaxies. This model makes several falsifiable predictions, including higher metallicities and dust masses than anticipated at high redshift and isotopic abundances in the Milky Way. It also predicts that stellar mass-to-light ratios are lower than produced using a Milky Way-derived IMF, so that inferences of stellar masses and star formation rates for high redshift galaxies are overestimated. In some cases, this may also transform inferred dark matter profiles from core-like to cusp-like.
\end{abstract}




\section{Introduction}

One of the most remarkable recent discoveries in galaxy evolution has been the surprising similarity in the properties of evolving galaxies on macroscopic scales, to the point that it has been possible to observationally describe the properties of a `normal' evolving galaxy.  It was already known that elliptical galaxies follow a series of scaling relations between radius, velocity dispersion, and surface brightness known as the `fundamental plane'~\citep{Gudehus1973,Pahre1998,Bernardi2003}.  More recently, star-forming galaxies at constant stellar mass and redshift have been shown to exhibit only a narrow range of star formation rates (SFR), a result that has come to be known as the star-forming main sequence~\citep{Brinchmann2004,Noeske2007,Peng2010,Speagle2014}.  There is a corresponding similarity for quasar accretion, relating luminosity to virial black hole mass and redshift~\citep{Steinhardt2010a,Steinhardt2011}.

This similarity is particularly surprising because, in modelling star formation and quasar accretion, it would seem that every process which affects gas availability, density, composition, dynamics, temperature, and pressure must be considered.  
Furthermore, spatial degrees of freedom could well be relevant, even though there have been successful modelling efforts which neglected these~\citep{Peng2010,Bouche2010,Dekel2014}.  

Although intuitively this complexity might be expected to lead to a chaotic solution in which every galaxy has an independent evolution driven by the local environment, attractor solutions (i.e., solutions in which different initial conditions lead to similar behaviour) could also exist~\citep{IKEA}.  Main sequence stars are locally chaotic, but feedback driving stars towards hydrostatic and thermal equilibrium results in an attractor solution for their macroscopic behaviour~\citep{Cox1968}.  Thus, it can be straightforward to predict the lifetime of the Sun although it is difficult to predict individual solar flares~\citep{Hathaway2015}.

We are therefore motivated to search for a similar equilibrium condition that might drive evolving galaxies toward a common path. 
In recent work, the authors considered the evolution of the stellar initial mass function (IMF) with temperature in star-forming clouds~\citep{Jermyn2018}, showing that a galactic IMF will be bottom-lighter than the Milky Way IMF when the temperature in star-forming regions exceeds the $\sim 20$ K of Milky Way star-forming clouds.  Indeed, both dust~\citep{Magdis2012,Casey2013,Magnelli2014,Magdis2017,Zavala2018} and gas~\citep{Daddi2015} temperatures in evolving galaxies are measured at $> 20$ K, implying that the IMF should be bottom-lighter than the Milky Way-derived IMFs commonly used in photometric template fitting of their properties~\citep{Arnouts1999,Ilbert2006,Conroy2009,Kriek2009}.  An increase in temperature also means that fewer molecular clouds can collapse and thus decreases the overall SFR, a process that has been at the heart of models describing feedback between the evolution of the central supermassive black hole and its host galaxy~\citep{Alexander2012,Kormendy2013,Somerville2015}.

Further, unlike the Milky Way, at high SFR, the dominant contribution to gas temperature likely comes from cosmic rays generated in the deaths of the most massive main sequence stars~\citep{Papadopoulos2010,Papadopoulos2012,Leite2017}.   Thus, cosmic rays provide a feedback loop: an increase in cosmic ray generation will decrease SFR and provide a bottom-lighter IMF, which in turn changes the rate which new cosmic rays are produced.  Depending upon the precise temperature-dependence of the SFR and IMF, both equilibrium and runaway solutions might be possible, as are more complex chaotic solutions.  

Considering cosmic rays as the primary mechanism for driving temperature is natural for two additional reasons.  Unlike photons and most other products of main sequence stars and stellar death, cosmic rays permeate gas throughout the entire galaxy, contributing to temperatures even in the coolest star-forming clouds.  Moreover, despite their likely importance, cosmic rays have proven difficult to include in current numerical simulations, so that a closer consideration of their properties is natural.  Indeed, their complexity has led to several other proposed cosmic ray feedback mechanisms, most notably cosmic winds \citep{Booth2013,Muratov2015,Hafen2019}

We outline the way in which star formation, the initial mass function, and cosmic rays interact in \S~\ref{sec:model}.
In \S~\ref{sec:equilibrium}, we show that this interaction results in stable attractor solutions at fixed gas mass. We then combine these interactions in \S~\ref{sec:simulations} with a basic numerical model of galaxy evolution which tracks the evolution of a halo, infalling gas, and a stellar population.  In some regimes the attractor is dominated by cosmic ray heating, and the properties of this track are compared with the star-forming main sequence in \S~\ref{sec:obs}.  Whether this idea can be developed into a reasonable toy model for galaxy evolution is considered in \S~\ref{sec:discussion}.

\section{Motivation}

We are motivated by three main considerations, each of which we will describe briefly below.  In combination, these suggest the development of the model proposed in this work.

\subsection{Temperature Dependence of Star Formation}

One of the most important processes governing star formation is the competition between gravity, driving the production of compact objects, and thermodynamics, which drives gas to expand rather than contract\footnote{Magnetic pressure and radiative effects also likely play a role, but our focus here is on the balance between temperature and gravity.}.  Therefore, a change in gas temperature will necessarily affect the balance between these two forces, changing the nature of galactic star formation.  This balance is often expressed in the framework of calculating a Jeans mass \citep{Jeans1902}, although there are several other effects which cannot be neglected in a full treatment.  

An increase in gas temperature will increase the Jeans mass, and therefore fewer molecular clouds will collapse.  Those which do collapse will also collapse more slowly.  As a result, galactic star formation rates (SFR) will decrease with increasing gas temperature.

At the same time, the clouds which do collapse at higher temperatures will also produce a different stellar mass distribution, since the fragmentation process is temperature-dependent \citep{Larson1985,Clarke2003,Jermyn2018}.  In~\citet{Jermyn2018}, we provide a model of this temperature dependence and argue that an increase in stellar mass temperature produces a bottom-lighter (or, equivalently, top-heavier) stellar initial mass function.  At increased temperature, galaxies will have a lower star formation rate, but the stars produced will be more massive, and in some cases this could even lead to an increase rather than a decrease in the production of O stars despite a lower overall SFR.

\subsection{High-redshift Galactic Temperatures}

Although gas in the Milky Way is typically around 20 K \citep{Papadopoulos2010}, observations of higher-redshift star-forming galaxies indicate that their gas is likely at higher temperatures.  Although it is difficult to measure the gas temperature directly, dust temperatures are more approachable.  \citet{Magnelli2014} find that dust temperatures along the star-forming main sequence range from 25 K to 40 K, with increasing temperature towards higher redshifts.  In addition, higher SFR at fixed redshift leads to a temperature increase, as might be expected due to the combination of higher luminosity and higher supernova rates.

It should be stressed that reasoning from luminosity-averaged dust temperatures to gas temperatures specifically in star-forming clouds introduces several major assumptions which are poorly justified.  It is very likely that star-forming clouds take on a variety of temperatures, just as observations of resolved galaxies indicate that star formation at any given time is typically concentrated only in some regions of star-forming galaxies (cf. \citet{Silverman2015}).  Nevertheless, the most reasonable interpretation of the available observational evidence is that star-forming galaxies sustain gas temperatures higher than those in the Milky Way.  

At $z \gtrsim 6$, the CMB temperature drives gas clouds to $T > 20\textrm{ K}$, and even if star-forming galaxies are hotter than the Milky Way, at some sufficiently high redshift, the CMB will dictate gas temperatures.  We consider this regime in building initial conditions for the model developed in \S~\ref{sec:model}, but there are currently no measurements of gas or dust temperatures at these extreme redshifts.

\subsection{Cosmic Ray Temperature Driving}

In the Milky Way, the $\sim 20\textrm{ K}$ typical temperature of cold gas in molecular clouds \citep{Elmegreen2008,Papadopoulos2011} is driven primarily by stellar radiation.  Cosmic rays alone would only drive the temperature to $\sim 5\textrm{ K}$ \citep{Goldsmith1978,Bergin2007,Papadopoulos2011}, so their contribution is negligible.  

However, at higher SFRs a corresponding increase in the supernova rate will result in a higher cosmic ray density.  A galaxy with an SFR of $100 \textrm{M}_\odot/\textrm{yr}$ will produce cosmic rays at 100 times the Milky Way rate.  The corresponding temperature increase is complex, with simulations estimating an average temperature of\footnote{ To obtain this form, it was assumed that the cosmic ray ionization rate per ${\rm H}_2$ molecule is directly proportional to $\rho_{\rm CR}$ in equation (12) of~\citet{Papadopoulos2010}. Following their estimates for the former, $5\times 10^{-17}{\rm s^{-1}}$ was assumed for the Milky Way.  For similar reasons, an ${\rm H}_2$ density of $10^4{\rm cm}^{-3}$ Milky Way ISM temperature of $T_{\rm Milky Way} \approx 20 {\rm K}$ were used.}
\begin{equation}
    \bar{T}_{\rm CR} \approx 0.32\left(\left(5\bar{n}^{1/2} \bar{\rho}_{\rm CR} + 0.28 \bar{n}^3\right)^{1/2} - 0.53\bar{n}^{3/2}\right)^{2/3},
\label{eq:tcr}
\end{equation}
\citep{Papadopoulos2010}, where $T_{\rm CR}$ is the contribution of cosmic rays to the temperature, $n$ is the density of ${\rm H}_2$ and $\rho_{\mathrm{CR}}$ is the average cosmic ray density.
The over-bar indicates normalisation with respect to the Milky Way, i.e., all over-barred temperatures are divided by the Milky Way ISM temperature of $20{\rm K}$.
At values that are likely typical for $z \sim 1-6$ star-forming galaxies, this will lead to a temperature dominated by cosmic rays rather than starlight or the CMB.

Critically, cosmic rays are predominantly produced by supernova shocks~\citep{Drury2012}.
Therefore, the details of the stellar population determine the rate at which cosmic rays are generated, while the bulk properties of the galaxy determine their transport.  Thus, the rate at which new cosmic rays are generated is in turn related to the current $\rho_{\rm CR}$.  

\section{A Cosmic Ray-Driven Model}
\label{sec:model} 

\begin{figure}
\includegraphics[width=0.47\textwidth]{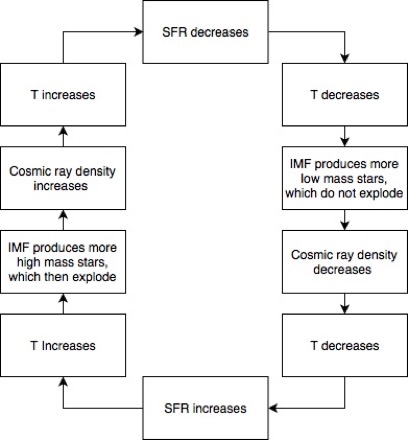}
\caption{The feedback cycle between star formation, the IMF, cosmic rays and galactic temperature.  In this work, we develop a model for galaxy evolution driven by this cycle and examine its properties.}
\label{fig:cycle}
\end{figure}

These three components suggest a model of the galactic star formation in which the CMB, cosmic rays and stellar radiation conspire to set the mean temperature gas clouds.  That temperature, in turn, determines the rate of star formation and the initial mass function, which both serve to set the density of cosmic rays and the intensity of stellar radiation.  These effects combine to form a feedback loop, which is depicted diagrammatically in Figure~\ref{fig:cycle}.

This loop is of particular interest because it directly relates the star formation rate and distribution at any point in time to the historic stellar population.
It is also comparatively simple yet potentially results in sufficiently complex behaviours as to be physically interesting.
Our aim is therefore to model and understand the phenomenology of this cycle.

Although the various components of this cycle are complex, there are relatively few observationally well-constrained properties of evolving galaxies.  At higher redshifts, these are generally averaged over the entire galaxy, even though higher-resolution and lower-redshift observations find that galaxies often have complex structure with different, e.g., star formation rates in different regions \citep{Nieten2006,Silverman2015}.  

As a result of these observational constraints, it is only possible to produce a meaningful model if there are very few parameters.  Thus, here we will treat the cycle as acting in a spatially homogeneous galaxy.  We further neglect all temporal variation in metallicity, and treat cosmic rays as being absorbed by the interstellar medium on a fixed time-scale.  Even with these considerable simplifications we require several additional ingredients.

\subsection{IMF}

We obtain the temperature-dependent IMF from~\citet{Jermyn2018}, who find
\begin{align}
\xi(M,\bar{T}) = \frac{dN}{dM}(\bar{T})\propto \begin{cases}
M^{-0.3}, & M < 0.08 M_\odot f(\bar{T}) \\
M^{-1.3}, & 0.08 M_\odot f(\bar{T}) < m < 0.50 M_\odot f(\bar{T})\\
M^{-2.3}, & 0.50 M_\odot f(\bar{T}) < m,
\end{cases}
\label{eq:newIMFfull}
\end{align}
where
\begin{align}
f(\bar{T}) = \bar{T}^2.
\end{align}
The proportionality factors in equation~\eqref{eq:newIMFfull} are such that $\xi$ is continuous and 
\begin{align}
    \int_{0}^{\infty} M\xi(M, \bar{T}) dM = 1 M_\odot.
\end{align}
Hence $\xi dM$ is the number of stars in a mass bin of width $dM$ given $1 M_\odot$ of star formation.
Note that this choice requires that the proportionality factors depend on $\bar{T}$.

\subsection{SFR}
\label{subsec:sfr}
Next we must determine the SFR as a function of temperature.
Note that the mass distribution of clumps in the Milky way is 
\begin{equation}
n(M) \propto M^{-1.60 \pm 0.06}
\end{equation}
where clumps are massive enough ($10M_\odot$ up to $3000M_\odot$) to allow measurement \citep{GomezWyrowski2014}.  We take this distribution to hold for all $M$.  It is further assumed (cf. \citet{Elmegreen2011} that there is negligible temperature dependence to this mass distribution.

Because Jeans timescale for collapse is independent of mass~\citep{Jeans1902}, this implies that the star formation rate at a given temperature is proportional to the number of clumps that exist above the direct collapse Jeans mass, $M_J \propto T^{3/2}$. Thus, we find that
\begin{equation}
\textrm{SFR} \propto \int_{M_J}^{\infty}{M^{-1.60 \pm 0.06} dM} \propto M_{\rm gas} T^{-0.90 \pm 0.09},
\label{eq:sfrt}
\end{equation}
where $M_{\rm gas}$ is the total gas mass of the galaxy.
Hence, scaling from the Milky Way, we obtain
\begin{align}
    \bar{\rm SFR} = \bar{M}_{\rm gas}\bar{T}^{-0.90\pm 0.09}.
    \label{eq:sfr}
\end{align}

\subsection{Cosmic Rays}

Once formed, stars of mass $M$ evolve and ultimately die over a timescale of
\begin{align}
    \tau \approx 10^{10} {\rm yr}\left(\frac{M}{M_\odot}\right)^{-5/2}.
\end{align}
The most massive of these new stars generate cosmic rays when they explode.
The overall cosmic ray generation rate in the Milky Way is well-approximated by assuming that each supernova converts an approximately fixed fraction $q \approx 1\%$ of its energy into cosmic rays~\citep{Smartt2009,Strong2010,Janka2012}.
This is in good agreement with theoretical models of supernova shocks~\citep{Drury1989} and so we assume that the cosmic ray generation from a single star is
\begin{equation}
	{\rm CRR}(M) = q E_\mathrm{SN}(M),
\end{equation}
where CRR is the cosmic ray generation rate and $E_\mathrm{SN}$ is the supernova energy associated with a star of mass $M$.

The majority of the galactic supernova energy budget arises from core-collapse and pair instability supernovae, with of order $E_{\rm CC/PI} \approx 10^{53}\mathrm{erg}$ emitted in each such event, somewhat independent of the progenitor mass~\citep{Kasen2011,Janka2012}.
By contrast, each Type Ia event emits of order $10^{51}\mathrm{erg}$~\citep{Mazzali2001}, and these do not make up a large enough fraction of supernovae to counterbalance their lower energy output~\citep{Cappellaro}.
As a result, the cosmic ray production is proportional to the number of stars which can undergo as supernova other than Type Ia, which is just those with masses greater than $8M_\odot$~\citep{Heger2003}.
Putting these pieces together we may write the total cosmic ray production as
\begin{align}
{\rm CRR}(t) = \frac{q E_{\rm CC/PI}}{M_\odot}\int_{8M_\odot}{\rm SFR}(t - \tau(M)) \xi(M,\bar{T}(t - \tau(M))) dM,
\end{align}
or
\begin{align}
    \bar{{\rm CRR}}(t) = \int_{8M_\odot}\bar{{\rm SFR}}(t - \tau(M)) \xi(M,\bar{T}(t - \tau(M))) dM.
\label{eq:CRRt}
\end{align}

In addition to being produced, cosmic rays are lost to a variety of processes.
Let $\tau_{\rm CR}$ be the characteristic loss time-scale.
Then
\begin{align}
    \frac{d \rho_{\rm CR}}{dt} ={\rm CRR}(t) - \frac{\rho_{\rm CR}}{\tau_{\rm CR}}.
    \label{eq:drho}
\end{align}
If we treat the Milky Way as being in cosmic ray steady-state and suppose that $\tau_{\rm CR}$ is fixed across galaxies then
\begin{align}
    {\rm CRR}_{\rm Milky Way} = \frac{\rho_{\rm CR, Milky Way}}{\tau_{\rm CR}},
\end{align}
so
\begin{align}
    \frac{d \bar{\rho}_{\rm CR}}{d(\tau_{\rm CR} t)} = \bar{{\rm CRR}}(t) - \bar{\rho}_{\rm CR},
    \label{eq:drho2}
\end{align}
which is solved by
\begin{align}
    \bar{\rho}_{\rm CR} = \int_0^{t} \bar{{\rm CRR}}(t')e^{-\tau_{\rm CR} (t-t')}  dt'.
    \label{eq:rhoCR}
\end{align}

\subsection{Gas Inflow}
\label{subsec:kolmogorov}
Although our toy model treats the mass distribution of cool gas clumps as fixed (\S~\ref{subsec:sfr}) this is only an approximation and so it is important to consider when and how it breaks down.
We begin with a Boltzmann-like transport equation for these clumps:
\begin{align}
    m\frac{\partial n(m)}{\partial t} = S(m) - \frac{\partial F}{\partial m},
    \label{eq:Boltzmann}
\end{align}
where $S(m)$ is a source of clumps of mass $m$ and $F$ is the rate at which clumps of mass $m$ fragment into clumps of mass $m-dm$.
In writing equation~\eqref{eq:Boltzmann} in this form we approximate fragmentation as a differential process in which $N$ clumps of mass $m$ become $N + N dm/m$ clumps of mass $m - dm$.
This is, however, consistent with the finding of simulations~\citep[][note that the merger rate in their equation~(1) diverges towards low masses]{Fakhouri2010} and is conceptually similar to the energy flow assumption underpinning the highly-successful Kolmogorov cascade~\citep{Kolmogorov41} so it is likely not wholly inappropriate.
Note that equation~\eqref{eq:Boltzmann} is only valid for $m \gg M_J$.
Near and below $M_J$ we expect clumps to either cease fragmenting or else form stars.
We therefore take the star formation rate to be of order $F(M_J)$.

The source term in equation~\eqref{eq:Boltzmann} is just gas inflow.
Because fresh gas tends to have large-scale structure we see that $S(m)$ has most of its support at large masses $m > M_{I}$, where $M_I$ is just the scale of the typical clump in the inflow.
We therefore have an equation with a source at large $m$ and a sink at $m > M_J$.
In steady-state, we would then have
\begin{align}
    F(M_J < m < M_{I}) = \int_{M_I}^{\infty} S(m') dm'.
\end{align}
If fragmentation is driven by gravitational interactions then above $M_J$ this process has no characteristic mass scale and hence is self-similar.
We therefore expect that
\begin{align}
    F(m) \approx \frac{1}{t_c} n(m) \left(\frac{m}{M_J}\right)^{\alpha},
\end{align}
where $\alpha$ is a dimensionless constant, $t_c = (G \rho)^{-1/2}$ is the gravitational free-fall time and $\rho$ is the mass density of the clump.
Because $F(m)$ is a constant in the intermediate regime we see immediately that $n(m) \propto m^{-\alpha}$, which is consistent with observations of a power-law distribution of clumps.

When the system is not in steady-state we might imagine perturbing $n(m) \rightarrow n(m) + dn(m)$ via a perturbation in $S(m)$.
The resulting perturbation propagates in $m$ so that
\begin{align}
    \frac{\partial(dn)}{\partial t} = -\frac{1}{t_c}\frac{\partial (dn m^\alpha)}{\partial m}.
\end{align}
If $\rho$ is approximately constant in $m$ then $t_c$ is as well, so the perturbation propagates with characteristic time $t_c$.
Hence our assumption in (\S~\ref{subsec:sfr}) that the distribution is in steady-state is good only so long as $S(m)$ varies slowly relative to the scale $t_c$.
For typical densities $t_c \approx 10^6 {\rm yr}$, which is quite short relative to the galaxy's dynamical timescale over which infall rates should vary, so our original assumption is justified.

\subsection{Temperature}

The final ingredient in this model is a relationship between the stellar population, the cosmic ray density, the CMB and the temperature.
Equation~\eqref{eq:tcr} gives the contribution of the cosmic ray density.
The temperature of the CMB is given by
\begin{align}
    \bar{T}_{\rm CMB} = 2.7 (1+z) \rm K,
    \label{eq:tcmb}
\end{align}
where $z$ is redshift.
Finally the contribution due to stellar radiation is
\begin{align}
    \bar{T}_{\rm rad} \approx \left(\frac{\bar{L}_\star}{\bar{M}_{\star}}\right)^{1/4},
    \label{eq:trad}
\end{align}
where $L_\star$ is the total luminosity of stars in the galaxy and $M_\star$ is the total stellar mass.
To combine these we let
\begin{align}
    \bar{T} = \max\left(\bar{T}_{\rm CR}, \bar{T}_{\rm rad}, \bar{T}_{\rm CMB}\right),
    \label{eq:tnet}
\end{align}
where $T_{\rm CR}$ is the temperature due to cosmic rays, $T_{\rm rad}$ is that due to stellar radiation and
This is a somewhat crude means of combining the two contributions, but it is accurate in regimes in which one effect dominates and provides a reasonable approximation even when this is not the case.

\subsection{Summary}

Equations~\eqref{eq:newIMFfull},~\eqref{eq:sfr},~\eqref{eq:CRRt},~\eqref{eq:drho},~\eqref{eq:tcr},~\eqref{eq:trad}, and~\eqref{eq:tnet} provide the quantitative specification of the feedback loop depicted in Figure~\ref{fig:cycle}.
Indeed together they comprise a minimal model of this feedback loop, neglecting spatial, chemical and magnetic degrees of freedom.  We might therefore hope that they could be understood analytically, searching for solutions with physical meaning.  As a system of multiple integrodifferential equations, these equations have multiple solutions and plenty of scope to support complex behaviour.  We consider the most physically interpretable of these in the following sections.

\section{Equilibrium}
\label{sec:equilibrium} 

The observed similarities between the growth of galaxies in different environments should motivate a search for equilibrium solutions.  As with the evolution of a main sequence star, a formal equilibrium is not required, but rather an attractor solution which slowly evolves as material is processed into stars would suffice. 

Thus, we will first search for solutions which would be in equilibrium holding gas availability fixed, then consider their broader time evolution on scales long enough that this approximation becomes invalid.  For any equilibrium solution, all of $T$, $\bar{\rho}_{\rm CR}$, and SFR are constant.

Holding $\bar{\rho}_{\rm CR}$ constant requires that
\begin{align}
    \overline{{\rm CRR}} = \bar{\rho}_{\rm CR}.
\end{align}
Inserting equation~\eqref{eq:CRRt} we find
\begin{align}
 \overline{\rho}_{\rm CR} = \int_{8M_\odot} \overline{{\rm SFR}} \xi(M,\bar{T}) dM,
\end{align}
where we have dropped the time dependence because we require the solution to be an equilibrium.
This further allows the SFR to be pulled out of the integral, so
\begin{align}
 \bar{\rho}_{\rm CR} = \bar{{\rm SFR}} \int_{8M_\odot} \xi(M,\bar{T}) dM.
\label{eq:equilib0}
\end{align}

When equation~\eqref{eq:tnet} is dominated by the CMB or stellar radiation equation~\eqref{eq:equilib0} is straightforward to solve because $\bar{\rho}_{\rm CR}$ does not affect any of the quantities on the right-hand side and so is a free parameter.
In this case equation~\eqref{eq:equilib0} is trivially satisfied, so the equilibrium condition reduces to
\begin{align}
    \bar{T} = {\rm constant.}
\end{align}
In the CMB-dominated case the temperature evolves according to the CMB and the stellar population is irrelevant.
In the stellar radiation-dominated case the equilibrium condition becomes
\begin{align}
    T = \bar{T}_{\rm rad} \approx \left(\frac{\bar{L}_\star}{\bar{M}_\star}\right)^{1/4}.
\end{align}
In equilibrium the IMF is constant, so there are $\xi(m,T) {\rm SFR} \tau(m)$ stars of mass $m$ present.
The contribution of stars of mass $m$ to $L_\star$ is therefore proportional to $\tau(m) L(m)$, or its lifetime energy output.
The specific energy output is approximately constant with stellar mass\footnote{It actually rises with stellar mass, but does so quite slowly and with finite non-zero asymptotic limits at both high and low mass~\citep{1989ApJ...347..998E}.} because $\tau \propto m^{-2.5}$ while $L \propto m^{3.5}$, so we expect a single equilibrium temperature in this limit, with temperature set by the spatial extent of the galaxy and the typical specific energy output of stars.
This solution is an attractor because the radiation temperature is set directly by the stellar population, and to leading order the stellar population has fixed specific energy output.

Finally, when cosmic rays dominate equation~\eqref{eq:tnet} we may expand equation~\eqref{eq:tcr} and approximate it by
\begin{align}
    \bar{T}_{\rm CR} \approx \frac{1}{2}\bar{\rho}_{\mathrm{CR}}^{1/3}.
\end{align}
Hence we find
\begin{align}
    \bar{T}^3 &= \frac{1}{8}\bar{\rho}_{\rm CR}\nonumber\\
    &= \frac{\bar{{\rm SFR}}}{8} \int_{8M_\odot} \xi(M,\bar{T}) dM.
\label{eq:equilib1}
\end{align}
As $\bar{T} \rightarrow 0$ the left-hand side vanishes, but the right-hand side  may be shown to scale as $T^{-0.3}$.
As $\bar{T} \rightarrow \infty$ the left-hand side diverges, while the right-hand side falls as $T^{-4.3}$.
By the racetrack theorem there must therefore be at least one crossover point between these limits, so there exists an equilibrium.
Indeed there is precisely one equilibrium, because the left-hand side is monotonically increasing while the right-hand side is monotonically decreasing.

The same argument shows that this equilibrium is also an attractor.  An input temperature above the main sequence solution results in less cosmic ray production than is required to maintain that temperature, and similarly a temperature below the main sequence solution will produce a temperature increase.

In the next section we provide numerical evidence for the existence of unique equilibria dominated by different temperature contributions and show that, excluding large fluctuations and crossovers between equilibria, they indeed appear stable.




\section{Numerical Solutions}
\label{sec:simulations}

The arguments in \S~\ref{sec:equilibrium} demonstrate that if the gas supply is held constant and the only non-negligible contribution to temperature is from cosmic rays, the system will tend towards an equilibrium solution with constant $T$ and SFR.  However, in practice neither of these is true.  

There is certainly long-term evolution in the gas supply, from a combination of processes such as star formation which use up available gas and mergers which increase the supply.  Therefore, the equilibrium towards which the system tends should be evolving.  Moreover, if the gas supply were to change on timescales shorter than the attractor timescale, it would be possible that the system would not tend towards a common evolution at all.  Similarly, the assumption that the gas temperature is dominated by the cosmic ray contribution might hold when a galaxy has a very high SFR, but should not apply to all galaxies or at all times.  These effects cannot be easily modelled with an analytical solution, but rather require solving the system numerically.

The full complexities of the problem are well beyond the simple model in these equations, and therefore although we track several relevant observables, the solutions presented here do not comprise a full simulation.  Rather, we merely examine the dynamics of solutions of our model in the presence of an evolving gas supply for star formation, an optional halo to supply new gas, and a mass-resolved stellar population.
These numerical studies allow us to both confirm the presence of attractor solutions and examine the long-term evolution of our model.  Although written in integrodifferential form,  the equations may also be cast as a set of ordinary differential equations specifying $T$, $\rho_{\rm CR}$ and the number of stars at each mass and time.  This discretised form is more convenient for numerical analysis and so is the one we use here.  

\subsection{Sample Galactic History}

\begin{figure}
\includegraphics[width=0.47\textwidth]{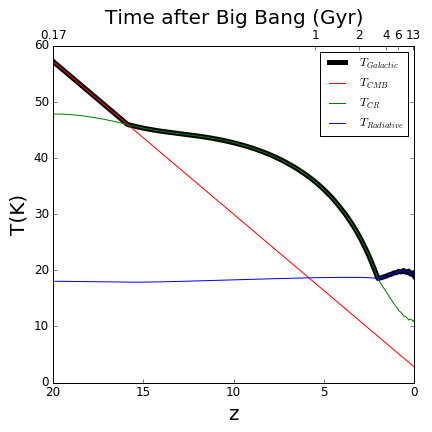}
\caption{The evolution of a galaxy starting at very high redshift with no stars and $10^{11} M_\odot$ of gas.  The history includes three phases: (1) a phase in which $T \sim T_{\textrm{CMB}}$ (red); (2) a phase in which $T$ is driven primarily by cosmic rays from star formation (green); and (3) a quiescent final phase in which $T$ is dominated by starlight from an existing, old stellar population (blue).  Although the cosmic ray contribution to temperature at high redshift is tiny, this is not due to a low SFR but rather a combination of the empty initial conditions and the substantial time delay between formation and explosion for all but the most massive stars.}
\label{fig:simresults}
\end{figure}

\begin{figure*}
\includegraphics[width=0.95\textwidth]{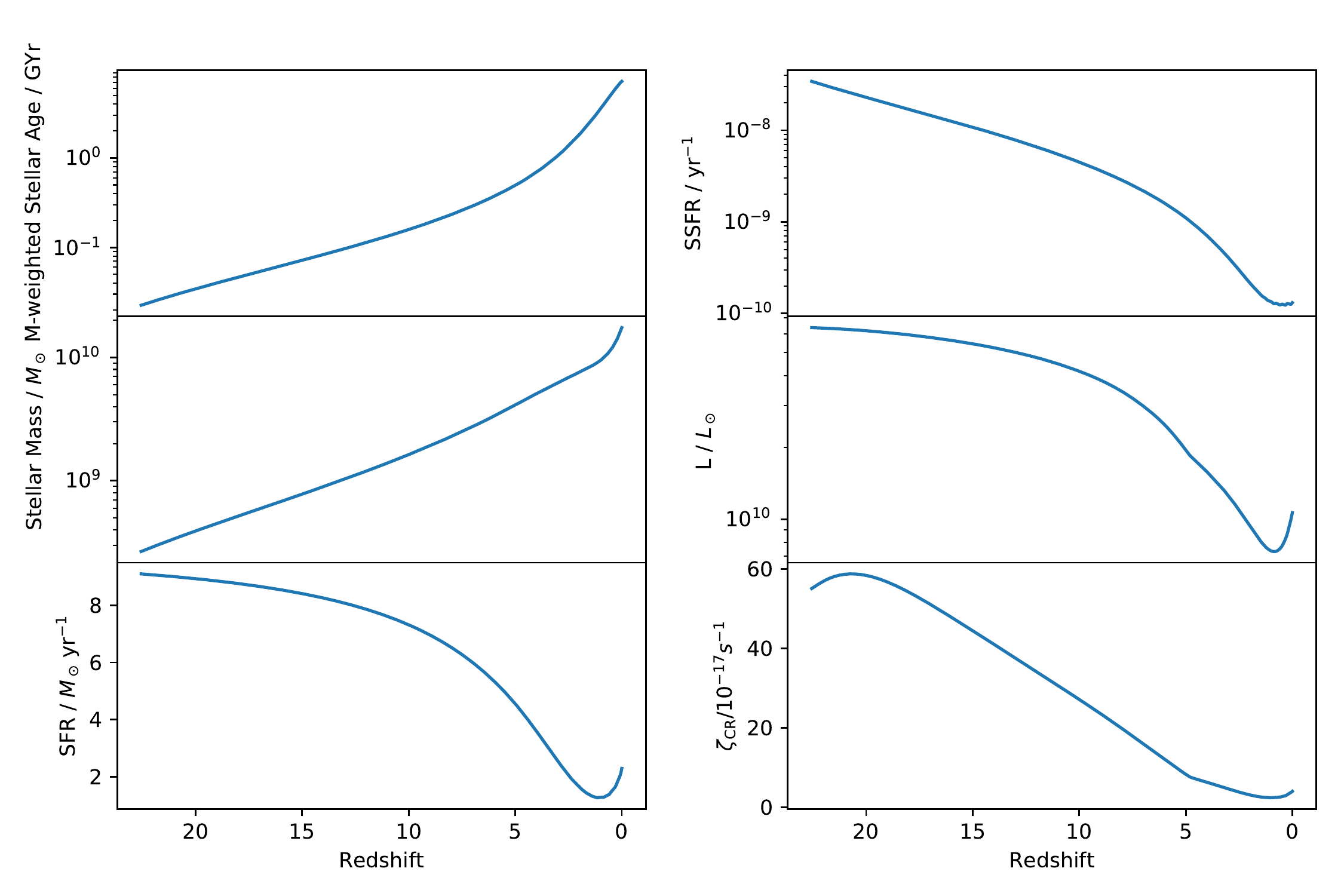}
\caption{The evolution of six observable parameters commonly fit for high-redshift galaxies as a function of redshift for a run which started with $10^{10}M_\odot$ of gas, no stars, and a $10^{11} M_\odot$ halo. Gas was added at a rate of $1.5\times 10^{-6}$ of the halo mass per year, and the halo mass was increased by a relative amount of $3\times 10^{-5}$ per year to simulate accretion.   The detailed evolution in each parameter can lie within a family of solutions, depending upon input parameters (e.g., Fig. \ref{fig:gasaddition}).  However, the overall trend in the evolution of each observable is more robust.}
\label{fig:outputs}
\end{figure*}

As a sample solution, consider the evolution of a galaxy which will eventually attain a stellar mass of $10^{11}$, similar to the Milky Way, from very high redshift until the present.
At early times the temperature will be dominated by the CMB.
Because this has an explicit redshift dependence, its include means the model is no longer a state machine, but now depends directly on cosmic epoch.
As the CMB cools cosmic rays and stellar radiation ought to come to dominate.
The result is an evolution that has three distinct phases (Fig. \ref{fig:simresults}). The corresponding evolution of several other parameters is shown in Fig. \ref{fig:outputs}.

At the highest redshifts, the temperature is dominated by the CMB~\citep[cf][]{Jermyn2018}.  The IMF at these temperatures is very top-heavy, so that few of these stars are present still at low-redshifts.
However, this phase results in a standardisation of galaxies: two halos containing the same baryonic mass but collapsing at different times will both enter this CMB-dominated phase and be attracted towards similar star formation rates and IMFs.  Because relatively few low-mass stars are made, this phase acts to destroy the history of when a galaxy collapsed.  Thus, by the time that the CMB has cooled to the point that it no longer dominates galactic temperatures, the two galaxies have ended up in very similar states despite their different origins.

As the CMB temperature declines, at some point the temperature becomes dominated by cosmic rays from supernovae.  This is the near-equilibrium, cosmic ray attractor solution described in \S~\ref{sec:simulations}.  With a fixed total gas supply, gas availability declines as stars are formed.  This is true even though much of the mass from supernova explosions produces metal-rich regions which can be cooled to form new stars, since the current SFR always exceeds the supernova rate from the relatively rare, high-mass stars formed in the recent past.  So, this solution produces a slow decline in SFR, along with a corresponding slow decrease in temperature and increasingly bottom-heavy IMF.  Most of even the low-mass stars in the galaxy are therefore formed at higher temperatures than at present, so that using a current Milky Way-derived IMF will overestimate the stellar mass even in the local universe.

In practice, there will additionally be gas inflow from mergers, likely initially dominated by hot gas but some of which will eventually become cool gas that can be processed into stars.  
\begin{figure}
\includegraphics[width=0.47\textwidth]{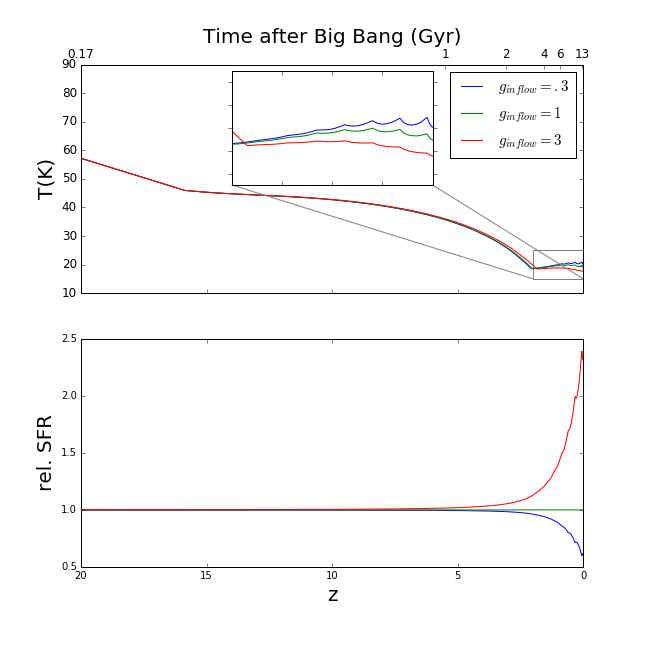}
\caption{The evolution of the same galaxy as in Fig. \ref{fig:simresults} with three different prescriptions for gas inflow.  The galaxy again starts at very high redshift with no stars and $10^{11} M_\odot$ of gas, and again goes through the same three phases, with similar behaviour at high redshift when there is a surplus of gas which cannot be processed into stars.  At late times, the details of gas inflow become more important, allowing star formation rates and the timing of the transition to starlight domination and quiescence to be tuned to match observations.  For the same reason, it is necessary to measure or constrain gas inflow rates to quantitatively predict the properties of main sequence rather than merely match it.}
\label{fig:gasaddition}
\end{figure}
The inflow of cool or cooling gas from minor mergers is poorly constrained by current simulations \citep{Nelson2015,Nelson2016}.  Halo merger rates are more robustly predicted (e.g., \citet{Fakhouri2010}).  However, even if the dark matter infall rate were well known, the baryon content, and in particular the gas fraction and stellar population of infalling material are unknown.  In Fig. \ref{fig:gasaddition}, evolution for several choices of infalling gas rates are illustrated, including a scenario in which infalling baryons are given the same stellar population and gas fraction as the current state of the larger galaxy.

Depending upon the gas infall rate, a family of attractor solutions can be produced.  As a result, it will be possible to at least broadly match observations of the star-forming main sequence with the correct choice of gas infall rate.  It should be noted that this is not necessarily a feature; the ability to always choose a gas infall rate to match observations means that doing so is not properly a prediction of this model.  To use this as a test, it would be necessary to also have not just measurements of overall infall rates, but also a classification of infalling material to determine how much additional gas is available and how infalling stars add to the stellar population.  However, this model does make several other predictions which might be falsifiable, as discussed in \S~\ref{sec:obs}.

Finally, at low redshift the SFR drops to the point that starlight provides a greater contribution to the temperature than cosmic rays.  Like with CMB domination, under these conditions the SFR will be suppressed below the equilibrium solution produced by cosmic rays alone.  Therefore, this might provide a mechanism by which galaxies with a substantial, old stellar population suppress star formation and remain quiescent.  This is also the solution which best describes the current state of the Milky Way.

\subsection{Convergence and Rounding}

To ensure convergence the simulations were run at a variety of time-steps.  For steps of $dt = 3\times 10^5{\rm yr}$ and below the results were nearly independent of step size, while above $10^6{\rm yr}$ the results were strongly dependent on step size.  This reflects the fact that the lifetimes of the massive stars which contribute the most to the cosmic ray density are of order $10^6{\rm yr}$ and so are under-resolved for step sizes longer than this.

Rounding does not present an issue for our purposes.  The equations have an attractor solution and do not depend sensitively on the exact state.  Therefore, it should be expected that small rounding errors will not build up over time, which is supported by our finding identical long-term numerical behaviour for slightly different initial conditions.  Over the course of $\sim 10^5$ steps we incur some rounding error at the relative level of $10^{-15}$ per step. The populations in bins vary by a few orders of magnitude over the course of the simulation, so absent amplification by the time evolution the maximum rounding-induced drift in any bin is of order one part in $10^7$.

\section{Observational Evidence and Tests}
\label{sec:obs}

The model developed here accomplishes our primary goal of producing an attractor solution for star formation rates characteristic of the star-forming main sequence.  However, because the SFR and several other details depend upon the gas inflow rate, which is poorly constrained, for the correct choice of parameters this model can be made consistent with a wide range of possible observed star-forming main sequences.  Thus, the qualitative existence of the star-forming main sequence is robust, but its quantitative details do not present falsifiable predictions.  

Further, because this model is intended to be illustrative rather than complete, there are several important effects which have not been included but which will have a non-negligible impact on quantitative details.  However, beyond the existence of a main sequence, the model makes several other predictions which are at least qualitatively robust.  These can be compared with current and near-future observational constraints in an attempt to evaluate whether this is a plausible attractor mechanism for galaxy evolution.

\subsection{Gas (and dust) temperatures}

A generic feature of our model is that the gas temperature ought to be a decreasing function of redshift. At early times this is because the CMB dominates the gas temperature, at intermediate times because the cosmic ray density is falling, and at late times because the stellar population ages and becomes more bottom-heavy.

While the gas temperature is difficult to directly observe, the dust temperature has been observed \citep{Magnelli2014} and agrees with this qualitative picture.  At any fixed redshift, dust temperatures along the star-forming main sequence are nearly constant.  At fixed mass, galaxies with star formation rates increasing (decreasing) above the main sequence increase (decrease) sharply in dust temperature, providing the feedback described here.  Further, the characteristic star-forming main sequence dust temperature declines from $\sim 35$ K to $\sim 25$ K from $z \sim 2$ to the present, in line with the evolution shown in Fig. \ref{fig:simresults}.

It should be noted that these dust temperatures are not necessarily the same as gas temperatures, because the two may not have sufficient time or cross-section to reach equilibrium.  Further, the nature of photometric measurements is such that these are luminosity-averaged dust temperatures, which may not match those in the coolest, star-forming regions of these galaxies.  Nevertheless, if gas temperatures in star-forming regions exhibit similar behaviour to dust temperatures, this would match the qualitative predictions of the cosmic ray feedback model.  Other models with a similar evolution in SFR can also produce similar behaviour, although an empirical fit to the main sequence alone is likely to produce an exponentially increasing SFR instead \citep{Peng2010,Steinhardt2017}.

It is still perhaps tempting to compare the predictions of the model presented here more quantitatively with dust temperatures.  However, dust temperatures are currently not sufficiently certain to provide such a test.  The inferred dust temperature depends strongly on the chosen value of effective emissivity $\beta_\textrm{eff}$ (cf. \citet{Magdis2011}).  Values ranging from 1 to 2.5 are common in the literature \citep{Hildebrand1983,Casey2012,Bianchi2013,Berta2016,Schreiber2018}.  As a result, estimates of dust temperature from these various groups can vary by $\gtrsim 10$ K for the same observations of the same galaxy due to modelling differences.  

\begin{figure}
\includegraphics[width=0.47\textwidth]{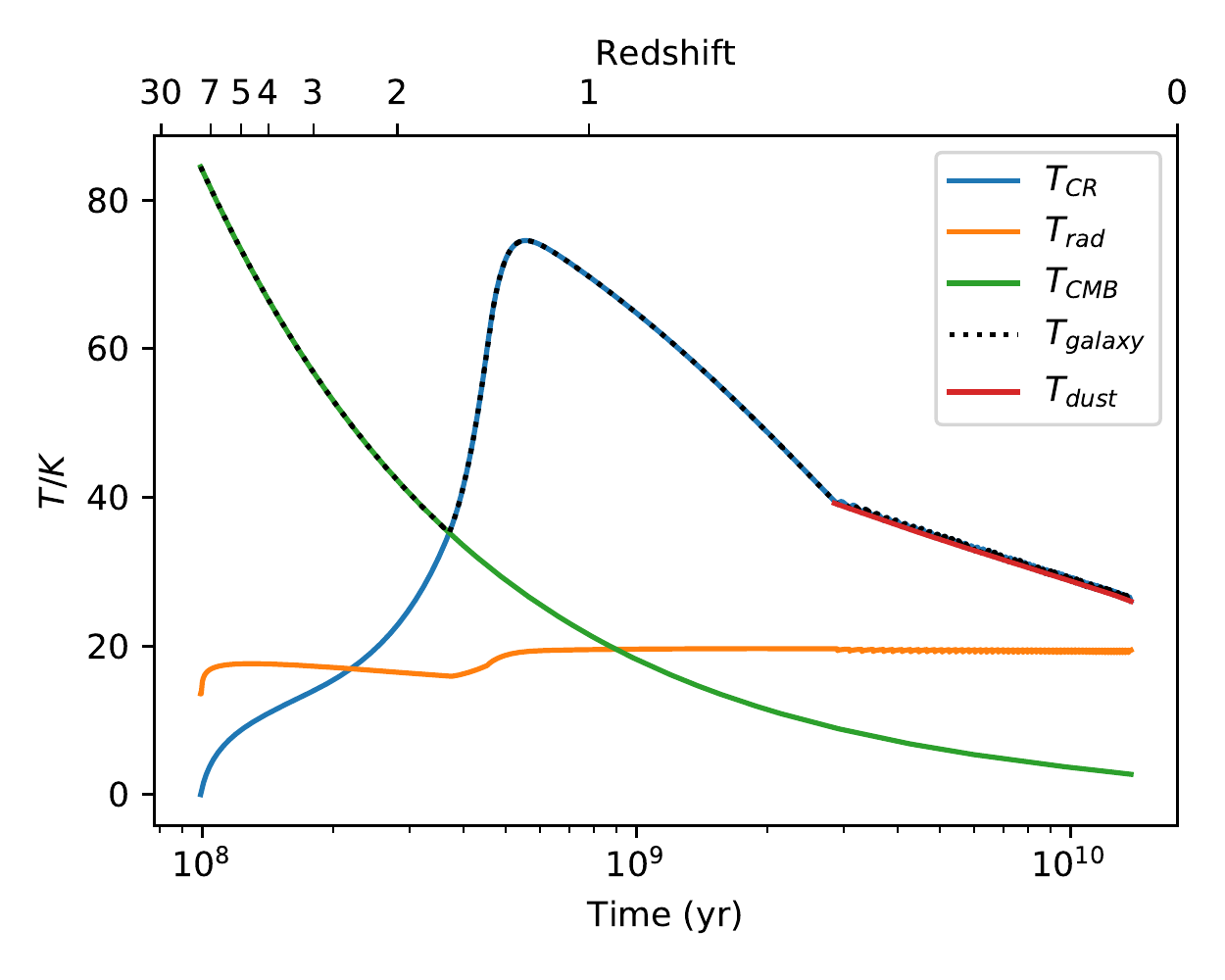}
\caption{The evolution of a galaxy with an initial gas mass of $10^9 M_\odot$ is shown for a gas infall rate tuned to match the temperature measurements reported by~\citet{Magnelli2014} in their equation (14) and the fit parameters specified therein. Good agreement is attained over the full reported redshift range, although this should not be taken as a test of our model because the gas infall rate could similarly have been be tuned to match a wide range of temperature observations.}
\label{fig:magnellimatch}
\end{figure}

We note that these different models do agree on general trends in dust temperature, which is why the trend in dust temperatures does provide a test of this model.  It should also be noted that because the cosmic ray-driven model does contain free parameters, most notably with respect to gas infall rates, it will be possible to tune the model (Fig. \ref{fig:gasaddition}) to empirically match the gas temperatures produced by any specific choice of $\beta_\textrm{eff}$ and other parameters.  For example, a closer match with the \citet{Magnelli2014} measurements is shown in Fig. \ref{fig:magnellimatch}.  However, because this arises from additional tuning, it cannot provide an additional test of the model presented in this work.

\subsection{High-redshift Metallicity and Dust}

A further feature of our model is that many massive, short-lived stars ought to be formed at high redshift and then almost immediately die.  This initially CMB-driven, and later cosmic ray-driven, phenomenon results in increased metal and dust production earlier than would otherwise be expected.  Because the total mass of the high-mass end of the IMF grows rapidly with temperature, and because at high redshift these temperatures are predicted to reach $\sim 40-75$ K rather than the $20$ K in the Milky Way, this can be a substantial effect, increasing dust and metallicity by a factor of up to 5-10.  

Recent observations have indeed found that the first galaxies are surprisingly dust- and metal-rich, to the point that it is difficult to produce dust \citep{Rowlands2014,Zavala2018} and possibly also metals \citep{Sparre2014,Hartoog2015} in such quantities without having had more star formation then currently believed.  The feedback described here therefore not only presents a possible solution to this problem but predicts that such a problem should exist.

\subsection{Chemical Abundances}

\begin{figure}
\includegraphics[width=0.47\textwidth]{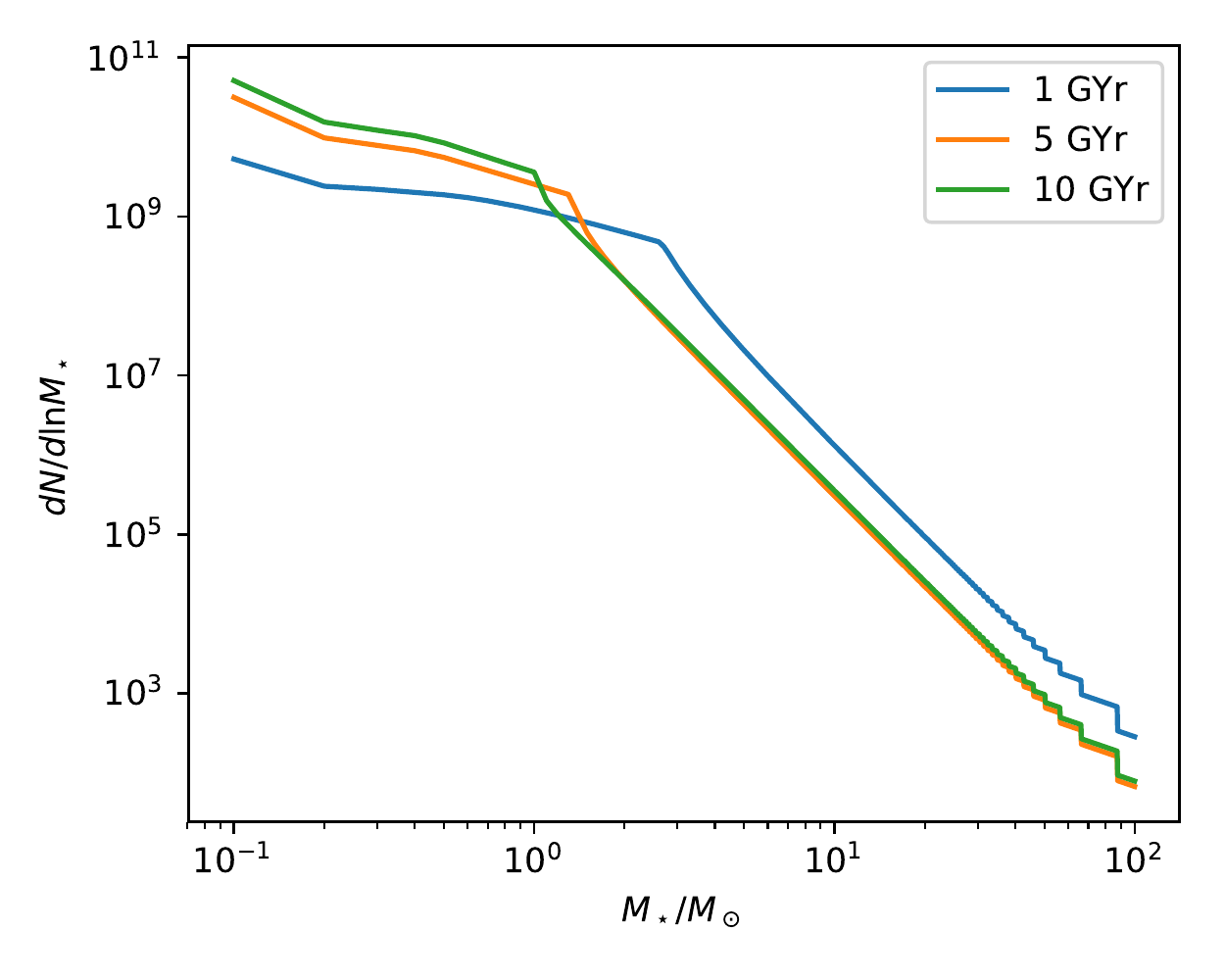}
\caption{The evolution of the stellar mass distribution for the same galaxy as in Fig.~\ref{fig:outputs}. At early times, the CMB dominates the IMF and forces a top-heavy stellar population. At late times, the gas is cold and therefore the mean stellar mass falls as the population ages.}
\label{fig:pop}
\end{figure}

The star formation history of a galaxy is a key factor in its chemical evolution.
As discussed by~\citet{Jermyn2018}, and illustrated in Fig.~\ref{fig:pop} at higher redshifts we predict higher typical stellar masses.
As a result there ought to be fewer low-mass asymptotic giants (AGB stars) at early times.
This should enhance $^{16}$O and other $\alpha$--process isotopes relative to carbon~\citep{1995ApJS...98..617T, Karakas2014}.
There may also be more $^{14}$N relative to $^{12}$C owing to more stars undergoing hot bottom burning~\citep{1995ApJS...98..617T,Karakas2014}.
Both effects have been seen in the Milky Way~\citep{2005ApJ...625..833L, 2012A&A...547A..76P}.

These effects ought to be strongest in galaxies which underwent a cosmic ray-dominated phase, and so should be stronger in more massive galaxies.
They should also be stronger in galaxies at in starburst galaxies, which may be in a runaway phase, as well as those at higher redshift.
Correspondingly they should be weaker in dwarf galaxies and in the local universe.
Hence while our model is too simple to predict the magnitude of these features, it does robustly predict the sign of their correlation with redshift, SFR, and galaxy mass.

\section{Discussion}
\label{sec:discussion}
 
This work develops a simplified model for galactic evolution based around feedback between gas temperature, star formation rate, and the stellar initial mass function.  For most of the history of a typical galaxy, the temperature is dominated by cosmic rays, which are perhaps the most important component currently not well modelled by numerical simulations.  It is therefore intriguing that considering cosmic rays nearly to the exclusion of all else produces, at least qualitatively, many of the features observed by large populations of star-forming galaxies.  

The most notable of these features is the prediction of two key features of the star-forming main sequence.  Cosmic ray feedback drives galaxies with similar gas mass but in different environments to a similar, attractor star formation rate.  Further, although this attractor solution is time-dependent, for proto-galaxies which produce enough metals to form main sequence stars sufficiently early (likely z$\gtrsim15$) that the initial temperature is dominated by the cosmic microwave background, the relevant time will be cosmic epoch rather than age since collapse.  Thus, galaxies do not evolve as a state machine, but rather with a dependence upon cosmic epoch, one of the most puzzling features of the star-forming main sequence.  

Thus, a cosmic ray-dominated model may be able to predict several of the behaviours that have been most difficult to produce in numerical simulations.  Further, the same model can describe a galaxy under very different conditions, as typical galactic history would include, in order, an initialisation phase, a star-forming main sequence phase, and a quiescent phase. However, a full treatment requires combining cosmic rays with the many other effects which are well-modelled by simulations, rather than the toy model presented here which excludes nearly everything else.  

Similarly, galaxies here are treated as monolithic, with one temperature, gas mass, cosmic ray density, and other properties.  This approximation is used both to make the problem tractable and because photometric template fitting typically produces only luminosity-averaged galaxy properties.  However, both observations and simulation demonstrate strongly that this is a poor model.  It is possible that the scatter in the main sequence, as well as conditions driving some galaxies to exhibit the runaway behaviour which is also a possible solution to these equations, is due to this effect.  However, a discussion of these solutions is reserved for a future companion paper.  

In short, it is intriguing that if one considers cosmic ray-driven evolution and excludes all of the other complex baryonic physics which simulations show to be important, the result is evolution which looks remarkably similar to observed features of evolving galaxies which have otherwise often been difficult to match.  However, we know that these other astrophysical effects are important, and we should not expect to produce a fully predictive model without including them.  It is hoped that presenting the mechanism in this format will motivate simulations which include these effects in combination with all of the others, despite the inherent difficulties.

\subsection{Impossibly Early Galaxies}

Recent high-redshift surveys have revealed a surprisingly large population of very early, very luminous, and therefore presumably massive, galaxies \citep{Steinhardt2014a,Song2016,Davidzon2017}.  Indeed, if these galaxies as truly as massive as has been derived from template fitting (out to $z\sim 6-8$) or UV luminosities ($z\gtrsim 8$), they are more massive than produced in simulations \citep{Genel2014} and should reside in halos too massive to have formed in the available time \citep{Steinhardt2016,Davidzon2017}.  Several different solutions have been proposed for this problem, including an abundance matching-derived shift in the stellar baryon fraction \citep{Finkelstein2015,Behroozi2015,Behroozi2018} and evolution in dust properties \citep{Mashian2016}.

This problem was a key motivation for our initial consideration of the relationship between temperature and the stellar initial mass function~\citep{Jermyn2018}, which showed that the CMB contribution to temperature in star-forming regions would decrease mass-to-light ratios, leading to stellar mass overestimation when using a Milky Way-derived IMF.  However, this effect is insufficient to reduce stellar masses to a locally-measured halo mass to stellar mass relation.  

The model described here predicts that the stellar populations in these galaxies formed at even higher temperatures, because galaxies at those masses were cosmic ray-dominated rather than CMB dominated even out to $z\sim 20$.  Therefore, mass-to-light ratios would be even lower, potentially to the point that stellar masses could be brought back into agreement with theory.  The large parameter space of our toy model prevents revised stellar masses from being a direct prediction of the model, but a sufficiently large correction is plausible.  Spectroscopy with JWST/NIRSpec will likely be sufficient to determine whether this is the correct resolution of the "impossibly early" galaxy problem.

\subsection{Dark Matter Distributions and Interactions}

Because massive galaxies become quiescent by the present epoch, most of their stars today are comprised of the low-mass end of the main sequences formed at higher redshifts.  The model developed here predicts that those main sequences formed at higher temperatures than present star-forming regions of the Milky Way.  Thus, again their stellar masses will be overestimated if a Milky Way-derived IMF is assumed.  For a typical evolutionary history, this results in a 20-40\% overestimate of stellar mass~\citep{Jermyn2018}.

Because stellar mass estimate is inherently difficult, for most applications, this correction is negligible.
However, it may shed light on a critical puzzle relating to the nature of dark matter.  The dark matter profile is inferred for local galaxies and clusters by subtracting the baryonic mass, which is inferred from stellar mass, from the total mass inferred from peculiar velocities (cf. \citet{Tulin2018}).  Recent observations find two populations: some galaxies exhibit a ``cuspy'' dark matter profile (\citet{Strigari2010,Breddels2013}; see also \citet{Genina2018}), in which the central density increases sharply as would be expected from collisionless, cold dark matter \citep{Navarro1996} and others exhibit a ``core'' with more constant central dark matter density \citep{Gentile2004,Gentile2007,deBlok2010}, which would require an additional interaction, either a dark matter self-interaction \citep{Elbert2015,Kaplinghat2016} or a dark matter-baryon interaction \citep{Robertson2018}.  Therefore, core profiles provide our strongest current evidence for dark matter interactions beyond $\Lambda$CDM.

At most radii, the dark matter density is higher than the baryon density.  For cuspy profiles, the same is true even in the centre of the galaxy, so that a 20-40\% overestimate in baryon mass is insignificant.  However, for core profiles, baryons become dominant in the central $\sim 1$ kpc in galaxies and a larger radius for clusters, in some cases attaining a maximum density 100 times higher than the constant dark matter density.  For such a galaxy, 20\% overestimate in the baryon density would mean that the dark matter density would not be constant, but rather would continue to increase, and will have been underestimated by as much as a factor of 20 in the centre.  This would be sufficient to turn a core profile into a cuspy profile, raising the possibility that with corrected stellar masses, no additional dark matter interactions would be required to match galaxy and cluster rotation curves.  

Whether the mass corrections would truly produce a cuspy profile as predicted for collisionless dark matter, or merely an increased central density, will depend upon model parameters.  It will also be sensitive to the additional effects described above which are beyond the scope of this work.  As with the other predictions of the toy model we develop here, these are only robust qualitatively, not quantitatively.  It should further be noted that in dwarf galaxies the stellar mass is always very low and, if the same is true of baryon densities, local dwarf galaxies with core profiles \citep{Zavala2013,Oh2015} will remain cores after this correction.


The simple feedback described here holds the possibility for producing many observed features of galaxy evolution which have heretofore been difficult to explain.
However its present incarnation is highly simplified and comes with a large permissible parameter space, which prevents us from making robust, quantitative predictions. 
Doing so will require fully integrating cosmic rays, in all of their complexity and difficulty, into numerical simulations which will be capable of making such predictions. 
Some progress on this front has already been made~\citep[cf][]{2018ApJ...868..108B}, and despite the inherent difficulty, the sharp differences between the properties of cosmic ray-dominated feedback and other models, as well as the problems which might potentially be solved, now make including these effects essential in future simulations.


The authors thank Iryna Butsky, Daniel Ceverino, Kristian Finlator, Vasily Kokorev, Adrian Lopez, and Georgios Magdis for helpful discussions.  CLS is supported by ERC grant 648179 "ConTExt".  The Cosmic Dawn Center (DAWN) is funded by the Danish National Research Foundation under grant No. 140.  ASJ thanks the UK Marshall commission and the Flatiron Institute of the Simons Foundation for financial support. Support for JL by the Rose Hills Foundation is appreciated. This research was supported in part by the National Science Foundation under Grant No. NSF PHY-1748958 and by the Gordon and Betty Moore Foundation through Grant GBMF7392.




\bibliographystyle{mnras}
\bibliography{ref}



\appendix


\bsp	
\label{lastpage}
\end{document}